\documentstyle[aps,twocolumn,epsfig,floats]{revtex}

\newcommand{\beq}{\begin{equation}}
\newcommand{\eeq}{\end{equation}}
\newcommand{\bdis}{\begin{displaymath}}
\newcommand{\edis}{\end{displaymath}}
\newcommand{\bea}{\begin{eqnarray}}
\newcommand{\eea}{\end{eqnarray}}
\newcommand{\barr}{\begin{array}}
\newcommand{\earr}{\end{array}}

\begin{document}
\draft
\wideabs{
\title{The universality class of absorbing phase transitions 
with a conserved field}
 
\author{Michela Rossi$^{(1,2)}$, Romualdo Pastor-Satorras$^{(2)}$,  
and Alessandro Vespignani$^{(2)}$}

\address{1) International School for Advanced Studies, SISSA/ISAS
  Via Beirut 2-4, 34014 Trieste, Italy\\
  2) The Abdus Salam International Centre 
  for Theoretical Physics (ICTP),
  P.O. Box 586, 34100 Trieste, Italy\\
}
\maketitle

\begin{abstract}
  We investigate the critical behavior of systems exhibiting a
  continuous absorbing phase transition in the presence of a conserved
  field coupled to the order parameter.  The results obtained point
  out the existence of a new universality class of nonequilibrium
  phase transitions that characterizes a vast set of systems including
  conserved threshold transfer processes and stochastic sandpile
  models.
\end{abstract}

}

Absorbing phase transitions (APT) are a category of critical
nonequilibrium phase transitions, widespread in condensed matter
physics and population and epidemics modeling \cite{reviews}.
Directed percolation (DP) \cite{reviews,kinzel83} has been recognized
as the paradigmatic example of a system exhibiting a transition from
an active to a unique absorbing phase.  DP defines a precise
universality class (theoretically described by the Reggeon field
theory \cite{grassberger79,rft}) which has proven to be very robust
with respect to the introduction of microscopic modifications.  The
Reggeon field theory is at the heart of a strong claim of
universality, summarized in the following conjecture
\cite{grassberger82}: {\em Continuous absorbing phase transitions to a
  unique absorbing state fall generically in the universality class of
  directed percolation}.  This conjecture is expected to hold for
models with short range interactions that, most importantly, do not
posses additional symmetries.

Many examples of APT subject to extra symmetries, and thus out of the
DP class, have been identified in recent years.  Among them we find
systems with symmetric absorbing states \cite{cardytauber96}, models
of epidemics with perfect immunization (the so-called dynamic
percolation class) \cite{dynperc}, and systems with an infinite number
of absorbing states \cite{many}.  Very recently, it has been
pointed out that the critical point of self-organized critical (SOC)
\cite{jenssen98,mannamodel} sandpile models can also be interpreted as
a continuous phase transition with many absorbing states
\cite{fes,bigfes}.  What distinguishes sandpile models from other
models with absorbing states, is that the control parameter,
represented by the global density of particles, is a conserved
quantity.
 
Given the large class of systems whose dynamics involves conserved
fields, it becomes particularly interesting to explore in general the
effect of conservation rules in APT. With this purpose in mind, in
this Letter we report the critical behavior of several models showing
absorbing transitions that strictly conserve the number of particles
or energy.  In particular, we introduce a conserved lattice gas (CLG)
\cite{noteje} with short range stochastic microscopic dynamical rules,
that undergoes a continuous phase transition to an absorbing state at
a critical value of the particle density.  We present extensive
numerical simulations in $d=2$ of the stationary and spreading
properties of the model, and determine the full set of critical
exponents. In order to prove definitively the existence of a
well-defined universality class we have also performed simulations of
a conserved threshold transfer process (CTTP) \cite{mendes94}, and
several fixed energy sandpile models with stochastic rules
\cite{mannamodel,fes,bigfes}.  All models provide critical exponents
compatible with a single and broad universality class that embraces
all APT in stochastic models with a conserved field. This evidence
leads us to conjecture that, in absence of additional symmetries, {\em
  absorbing phase transitions in stochastic models with infinite
  absorbing states and activity coupled to a static conserved field
  define a unique and per se universality class} \cite{notewij}.  This
result is relevant in the understanding of several reaction-diffusion
systems, sandpile models and activated processes that could share the
same theoretical description.

The CLG model is defined on a $d$-dimensional square lattice. To each
site $i$ it is associated a binary variable $n_i$ that assumes the
values $n_i=1$ if the site is occupied by a particle or $n_i=0$ if the
site is empty.  Double occupancy is strictly forbidden. Nearest
neighbors particles repel each other via repulsive short range
interactions. As a product of this interaction, at each time step
particles with nearest neighbors jump into one of their empty nearest
neighbor sites, selected at random.  The only dynamics in the model is
due to these {\em active} particles; isolated particles do not move.
The dynamics can be implemented with either sequential or parallel
updating.  In the latter case, an exclusion principle is applied so
that two particles never attempt to move into the same site.  We
impose periodic boundary conditions, and since the dynamics admits
neither input nor loss, the total number of particles $N=\sum_i n_i(t)$
is a conserved quantity. It is clear that the model allows an infinite
number (in the thermodynamic limit) of absorbing configurations, in
which there are no nearest neighbor particles.

\begin{figure}[t]


    \centerline{\epsfig{file=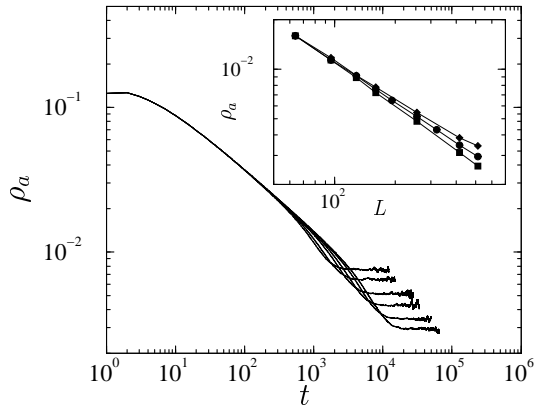, width=7cm}}
  \caption{Active-site density in surviving trials at the critical
    point for the CLG. From top to bottom, system sizes are $L=160,
    192, 256, 320, 416$, and $512$. Inset: Stationarity active-site
    density as a function of $L$ for a critical (center), subcritical
    (down), and supercritial (up) particle density.}
  \label{fig:steady1}

\end{figure}

\begin{figure}[t]

  \centerline{\epsfig{file=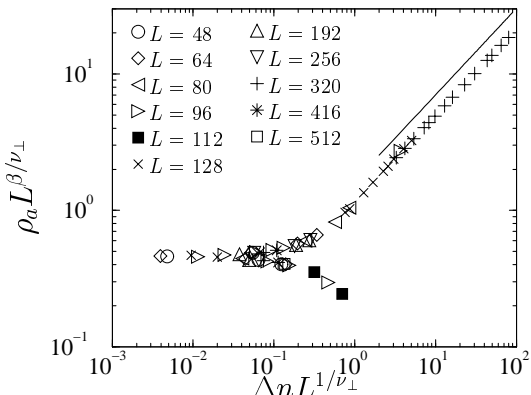, width=7cm}}
  
  \caption{Collapse of the stationary density  $\rho_a(\Delta n,
    L)L^{\beta/\nu_{\perp}}$  as a function of $\Delta n L^{1/\nu_{\perp}}$ for the
    CLG. The slope of the line is $0.63$.}
  \label{fig:steady2}
\end{figure}

In the CLG model, the constant particle density $n=N/L^d$ acts as a
tuning parameter. Initial conditions are generated by placing at
random in the lattice $nL^d$ particles, generating an homogeneous and
uncorrelated distribution.  For small densities, the system will very
likely fall into an absorbing configurations with only isolated
particles. For large densities, the system reaches a stationary active
state with everlasting activity (this is trivially the case for
$n>1/2$). We shall see in the following that as we vary $n$, the CLG
model exhibits a continuous transition separating an absorbing phase
from an active phase. The phase transition occurs for a nontrivial
density $n_c$ ($<1/2$).  APT are characterized by the order parameter
$\rho_a$ measuring the density of dynamical entities, in our case the
density of nearest neighbor particles. The order parameter is null for
$n<n_c$, and follows a power law $\rho_a\sim (n-n_c)^\beta$ for
$n>n_c$.  The system correlation length $\xi$ and time $\tau$ both
diverge as $n\to n_c^+$. In the critical region the system is
characterized by power law behavior, namely $\xi\sim
(n-n_c)^{-\nu_\perp}$ and $\tau\sim (n-n_c)^{-\nu_\parallel}$. The
dynamical critical exponent is defined as $\tau\sim\xi^z$, with
$z=\nu_\parallel/\nu_\perp$.  These exponents fully define the
critical behavior of the stationary state of the model.

In order to study the critical point of the CLG model, we performed
numerical simulations in $d=2$ for systems with size ranging from
$L=64$ to $L=512$, averaging over $10^4-10^5$ independent initial
configurations. Very close to the critical point we have $\xi\gg L$,
so that the actual characteristic length of the system is the lattice
size $L$. Because of its finite size, the system will enter sometimes
an absorbing configuration even for values of $n$ in the supercritical
region. It is then convenient to introduce averages over a set of
independent trials and calculate the quasi-stationary properties in
the active phase from a restricted average over surviving trials with
nonzero final activity.

As shown in Fig.~\ref{fig:steady1}, after a transient which depends on
the system size $L$ and $\Delta n\equiv n-n_c$, the surviving samples 
average of
the density of active sites reaches a stationary state $\rho_a(L,\Delta n)$.
Close to the critical point, the finite size scaling ansatz tells us
that all quantities depend on the system size through the ratio
$L/\xi$, and the order parameter follows the finite size scaling form
\cite{jensen93b}
\begin{equation}
  \rho_a (\Delta n,L) = L^{-\beta/\nu_{\perp}}  {\cal G} (L^{1/\nu_{\perp}} \Delta n) \;,
  \label{fss}
\end{equation}
where ${\cal G}$ is a scaling function with ${\cal G}(x) \sim x^{\beta}$ for
large $x$.  For $\Delta n=0$ the stationary density follows the pure power
law behavior $\rho_a\sim L^{-\beta/\nu_{\perp}}$. On the other hand, for values of $n$
in the supercritical regime $\rho_a$ should be independent of $L$ for
$L\gg\xi$, while in the subcritical regime $\rho_a$ should decay faster than
a power law. This allows us to locate the critical value $n_c$ of the
particle density as the only value of $n$ at which we recover a
nontrivial power law scaling for the density of active sites. In Fig.
~\ref{fig:steady1} we observe power law scaling for $n=0.23875$, but
clearly not for $0.2387$ or $0.2388$, indicating that
$n_c=0.23875(5)$ (Figures in parenthesis indicate the statistical 
uncertainty in the last digit). 
From the power law decay we find the exponent ratio
$\beta/\nu_\perp=0.81(3)$. An independent estimate 
of the exponent $\beta$ can be
obtained by looking at the scaling of the active-site density with
respect to $\Delta n$ for the size $L=320$. The
resulting power law behavior yields $\beta=0.63(1)$, where the error is
mainly due to the uncertainty in the critical point $n_c$.  A
consistency test can be performed by considering the active site
density away from the critical point. In Fig.~\ref{fig:steady2} we
plot $\rho_a(\Delta n, L)L^{\beta/\nu_{\perp}}$ 
versus $\Delta n L^{1/\nu_{\perp}}$ for
$\nu_{\perp}=0.78$, $\beta/\nu_{\perp}=0.81$ and $n_c=0.23875$. 
As one would expect
all the data collapse onto a single curve, following the scaling form
Eq.~(\ref{fss}). A further check is provided by the direct fitting
of the large $x$ behavior of the scaling function ${\cal G}(x)$ that
gives $\beta=0.63$, recovering the independent measurement at $L=320$.

To determine the dynamical exponents we turn our attention to the
scaling properties of time dependent quantities. In particular, we can
define a characteristic time by studying the decay of the probability
$P(t)$ that a random initial configuration has survived up to time $t$. At
the critical point this probability decays, at large times, as
$P(t)\sim\exp (-t/\tau)$.  At $\Delta n=0$ the effective 
characteristic length is
the system size $L$, and we have that $\tau(L)\sim L^z$. We can access the
value of $\tau(L)$ by a direct fitting of the $P(t)$ exponential tail;
$z$ is then estimated from the behavior of $\tau(L)$ for different $L$.
Again, a clean power law behavior is obtained for $n_c=0.23875$,
yielding $z=1.52(6)$.  Also in this case a consistency check can be
performed by studying the time decay of the active sites density
$\rho_{a,all}(t)$, averaging over all trials, even those that have
reached an absorbing state. Assuming a single characteristic time
scaling as $L^z$, we have at $\Delta n=0$ \cite{jensen93b}
\begin{equation}
  \rho_{a,all} (t,L) =   t^{-\theta}  {\cal F} (tL^{-z}) \;,
  \label{fsstime}
\end{equation}
where ${\cal F}(x)$ is a constant for $x\ll1$, and decays faster than
any power law for $x\gg1$. Data from simulations with different $L$ can
be collapsed onto a universal curve by plotting $\rho_{a,all} (t,L)t^{\theta}$
versus $tL^{-z}$. The best collapse is obtained for $\theta=0.43$ and
$z=1.52$, confirming the value obtained for the dynamical critical
exponent.  The exponent $\theta=0.43(1)$ is recovered also from a direct
fitting of the decay of the stationary density averaged over surviving
trials (see Fig.~\ref{fig:steady1}).  In usual APT, the latter
exponent obeys $\theta=\beta/\nu_\parallel$. This relation assumes a standard scaling
behavior at $\Delta n=0$ for $\rho_a(t)$.  In our model, however, the simple
scaling behavior is broken by an anomalous scaling regime (visible in
Fig.~\ref{fig:steady1} by the sharp drop just before the stationary
state) that seems to grow steeper with increasing $L$.  It follows
that data collapse in time is not achievable with standard scaling
forms, and that $\theta$ violates the usual scaling relation.  Albeit its
origin is not clear, it is noteworthy that this anomaly is common to
all APT with conserved fields inspected so far \cite{fes,bigfes}. At
this respect, it is interesting to note that the exponents of the
model fulfill all scaling relation in standard APT. In what respects
to hyperscaling relations, some of them, as $D=d + z - \beta/\nu_\perp$, are
fulfilled, while others, like $\eta+\delta+\theta=d/z$, are not. This is again due
to the $\theta$ exponent anomaly.

In APT it is possible to obtain more information
on the critical properties by studying the evolution (spread) of
activity in systems which start close to an absorbing configuration
\cite{grassberger79}. In each {\em spreading} simulation, a small
perturbation  is added to an absorbing
configuration.  It is then possible to measure the spatially
integrated activity $N(t)$, averaged over all runs, and the survival
probability $P(t)$ of the activity after $t$ time steps.  Only at the
critical point we have power law behavior for these magnitudes.

Here, we will follow a procedure equivalent to the definition of
slowly driven simulations in sandpiles, that enlights the
connections with these models. Instead of fixing the density $n$ by
working with periodic boundary conditions, and thus studying the
system at a given distance below the critical point, we impose open
boundary conditions and start each spreading experiment by adding a
new particle. Under these conditions, the system flows to a stationary
state with balance between the input of particles and
the boundary dissipation. In the limit in which the particle addition
is infinitely slow with respect to the spreading of activity, the
system reaches a critical state with density $n_c$ (in the
thermodynamic limit)\cite{braz}. The infinitely slow drive is implemented by
adding a new active particle only when the system falls into an absorbing
configuration. The system thus jumps between absorbing states via
avalanche-like rearrangements, and we can associate each
spreading experiment with an avalanche. 
The probability distribution $P_s(s)$ of having a spreading event 
involving $s$ sites, as well as the the
quantities $N(t)$ and $P(t)$ can be measured.  
The only characteristic length is the system size $L$,
 and we can write the scaling forms
\cite{grassberger79}:
\begin{eqnarray}
  N(t)&=&t^\eta f(t/L^z), ~~P(t)=t^{-\delta}g(t/L^z), \nonumber \\
  P_s(s)&=&s^{-\tau_s}h(s/L^D),
  \label{spread}
\end{eqnarray}
where the scaling functions $f(x)$, $g(x)$ and $h(x)$ are decreasing
exponentially for $x\gg1$, and we have considered that the spreading
characteristic time and size are scaling as $L^z$ and $L^D$,
respectively. Simulations were performed for system of size between
$L=64$ and $L=1024$, averaging over at least $5\times10^6$ spreading
experiments. The extrapolation of the measured densities at infinite
$L$ yields a critical density $n_c=0.2388(1)$, in perfect
agreement with steady state simulations. The scaling exponents are
measured using the now standard moment analysis technique
\cite{moments,mannasim}.  The resulting exponents are summarized in Table I.
In particular, the dynamical exponent
$z=1.53(2)$ is in excellent agreement with the stationary state
simulations, confirming the presence of a single critical behavior for
both cases.

\begin{table}[t]
\begin{tabular}{lccccc}
  &  \multicolumn{5}{c}{Steady state exponents} \\
  \cline{2-6}
  & $\beta$ & $\beta/\nu_\perp$ & $z$ & $\theta$ &\\
  \hline
  CLG &   $0.63(1)$ & $0.81(3)$ & $1.52(6)$ &$0.43(1)$&\\
  CTTP &$0.64(1)$  &$0.78(3)$  & $1.55(5)$ &$0.43(1)$& \\
  Manna & $0.64(1)$ & $0.78(2)$ & $1.57(4)$ &$0.42(1)$& \\ 
    DP & $0.583(4)$ & $0.80(1)$ & $1.766(2)$ & $0.451(1)$& \\
  \hline \hline
  &  \multicolumn{5}{c}{Spreading exponents} \\
  \cline{2-6}
  & $\tau_s$  & $D$ & $z$ & $\eta$  &  $\delta$ \\
  \hline
  CLG   & $1.29(1)$ & $2.75(1)$ & $1.53(2)$ & $0.29(1)$ &
  $0.49(1)$ \\
  CTTP   & $1.28(1)$ & $2.76(1)$ & $1.54(1)$ & $0.30(3)$ &
  $0.49(1)$ \\
  Manna & $1.28(1)$ & $2.76(1)$ & $1.55(1)$ & $0.30(3)$
  & $0.48(2)$ \\  
    DP    & $1.268(1)$ & $2.968(1)$ & $1.766(2)$ & $0.230(1)$ &
  $0.451(1)$ 
\end{tabular}
\caption{Critical exponents for spre\-ading and ste\-ady state
  experiments. Figures in parenthesis indicate the statistical 
  uncertainty in the last digit.   Steady state Manna exponents from
  Ref.~\protect\cite{bigfes}. }
\label{table}
\end{table}

\begin{figure}[t]

  \centerline{\epsfig{file=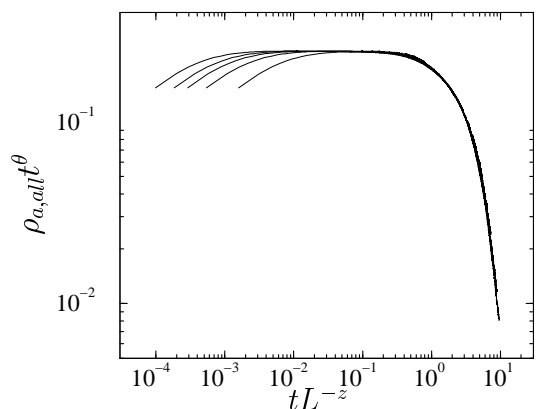, width=7cm}}
  
  \caption{Collapse of the active-site  density $\rho_{a,all}
    (t,L)t^{\theta}$  
    as a function of  $tL^{-z}$ for the CTTP. System sizes range from
    $L=64$ 
    to $L=384$.}
  \label{fig:steady3}
\end{figure}

In order to provide further evidence for the existence of a general
universality class, we have simulated several other models exhibiting
an APT in the presence of a conserved field.  The first is a conserved
threshold transfer process (CTTP). In the CTTP, the sites of a lattice
can be vacant, singly occupied, or doubly occupied by particles,
corresponding to a dynamic variable $n_i=0,1$ or $2$,
respectively. Values $n_i>2$ are strictly forbidden. Dynamics affects
only doubly occupied sites: every site with $n_i=2$ tries to transfer
both its particles to randomly selected nearest neighbors with
$n_j<2$. Singly occupied sites are, on the other hand, inert. The total
number of particles $N=\sum_i n_i$ is thus constant in time. Results
from simulations are obtained along the lines shown for the CLG and
are reported in Table~\ref{table}; in this case, the largest sizes
used are $L=512$ for the stationary exponents and $L=1024$ for the
spreading exponents.  As an example of our simulations, in
Fig.~\ref{fig:steady3} we plot $\rho_{a,all} (t,L)t^{\theta}$ as a
function of $tL^{-z}$, which shows a remarkable data collapse.  We
have also investigated the Manna sandpile model, and its variations by
the inclusion of a stochastic threshold\cite{ricepile}.  In this case,
an absorbing phase transition is obtained by using periodic boundary
conditions and a fixed number of sand grains (energy) as reported in
\cite{fes,bigfes}.   All models lead invariably to the
same universality class as the Manna model\cite{notabtw} (complete
results on these models will be reported elsewhere).

Our results provide striking evidence for a common critical behavior
which is incompatible with the DP universality class (see
Table~\ref{table})\cite{nota1dim}.  More noticeably, the models share
also the same scaling anomaly in the exponent $\theta$ \cite{bigfes},
signalling a common behavior in the transient regime to the stationary
state in the active phase. This uniformity of results confirms the
hypothesis of a unique universality class for all models with the same
conservation symmetry, and lead us to conjecture that, in the absence
of additional symmetries, {\em absorbing phase transitions in systems
with stochastic dynamics in which the order parameter is locally
coupled to a static conserved field define a single and new
universality class}.  This conjecture is supported by noticing that
the sandpile model has the same structure and basic symmetries of the
present CLG, once the field $n(x,t)$ is replaced by local energy (sand
grains) field \cite{fes,bigfes}. Indeed, in all models presented here,
a conserved non-critical field is dynamically coupled to the
non-conserved order parameter field $\rho_a$. Very likely this basic
structure will be reflected in a unique theoretical description (a
field theory with the same relevant terms and symmetries) that
accounts for the shared critical properties of these models.

This work has been supported by the European Network under Contract
No.  ERBFMRXCT980183. We thank D. Dhar, R. Dickman, P. Grassberger,
M. A. Mu{\~n}oz, and S. Zapperi for helpful comments and discussions.


\vspace*{-0.30cm}
\begin{thebibliography}{10}
\vspace*{-1.30cm}

\bibitem{reviews}
J. Marro and R. Dickman, {\em Nonequilibrium phase transitions in lattice
  models} (Cambridge University Press, Cambridge, 1999).

\bibitem{kinzel83}
W. Kinzel,  in {\em Percolation Structures and Process}, Vol.~5 of {\em Annals
  of the Israel Physical Society}, edited by G. Deutscher, R. Zallen, and J.
  Adler (Adam Hilger, Bristol, 1983), Chap.~18.

\bibitem{grassberger79}
P. Grassberger and A. {de la Torre}, Ann. Phys. (N.Y.) {\bf 122},  373  (1979).

\bibitem{rft}
J.~L. Cardy and R.~L. Sugar, J. Phys. A: Math. Gen. {\bf 13},  L423  (1980);
H.~K. Janssen, Z. Phys. B {\bf 42},  151  (1981).

\bibitem{grassberger82}
P. Grassberger, Z. Phys. B {\bf 47},  365  (1982).

\bibitem{cardytauber96}
J.~L. Cardy and U. T\"{a}uber, Phys. Rev. Lett. {\bf 77},  4780  (1996).

\bibitem{dynperc} 
  P. Grassberger, Math. Biosci. {\bf 63}, 157 (1982);
  H.~K. Janssen, Z. Phys. B {\bf 58}, 311 (1985).

\bibitem{many}
E.~V. Albano, J. Phys. A: Math. Gen. {\bf 25},  2557  (1992);
I. Jensen, Phys. Rev. Lett. {\bf 70},  1465  (1993).

\bibitem{jenssen98}
P. Bak, C. Tang and  K. Wiesenfeld, Phys. Rev. Lett. {\bf 59},  381 (1987);
For a review see: 
H.~J. Jensen, {\em Self-Organized Criticality} (Cambridge University Press,
  Cambridge, 1998).

\bibitem{mannamodel}
S.~S. Manna, J. Phys. A {\bf 24},  L363  (1991);
D. Dhar, Physica A {\bf 263},  4  (1999).

\bibitem{fes}
R. Dickman, A. Vespignani, and S. Zapperi, Phys. Rev. E {\bf 57},  5095
  (1998);
A. Vespignani, R. Dickman, M.~A. Mu{\~n}oz, and S. Zapperi, Phys. Rev. Lett.
  {\bf 81},  5676  (1998).

\bibitem{bigfes}
  A. Vespignani, R. Dickman, M.~A. Mu{\~n}oz, and S. Zapperi,
  cond-mat/0003285. 

\bibitem{noteje}
        The CLG can be considered the stochastic variant of a 
        lattice gas introduced 
        in the study of  $1/f$ noise in driven systems:
        H.J. Jensen, Phys. Rev. Lett {\bf 64}, 3103 (1990).

\bibitem{mendes94} 
        The CTTP is the conserved counterpart of the threshold
        transfer process introduced by J. F. F. Mendes, R. Dickman,
        M. Henkel, and M. C. Marques, J. Phys. A:  
        Math. Gen., {\bf 27} 3019 (1994).

\bibitem{notewij} Absorbing phase transitions coupled to a diffusive
conserved field belong to a different universality class: F. van
Wijland, K. Oerding, and H. J. Hilhorst, Physica A {\bf 251}, 179
(1998).
 
\bibitem{jensen93b}
        I. Jensen and R. Dickman, Phys. Rev. E {\bf 48},  1710  (1993).

\bibitem{braz}
        R. Dickman, M.~A. Mu{\~n}oz, A. Vespignani, and S. Zapperi, 
        cond-mat/9910454.

\bibitem{moments} 
  M. {De Menech}, A.~L. Stella, and C. Tebaldi, Phys.  Rev. E {\bf
    58}, R2677 (1998); C. Tebaldi, M. {De Menech}, and A.~L.  Stella,
  Phys. Rev. Lett {\bf 83}, 3952 (1999);

\bibitem{mannasim}
  A. Chessa, A. Vespignani, and  S. Zapperi, Comput. Phys. Comm. {\bf
    121-122}, 299 (1999);
  S. L{\"u}beck, Phys. Rev. E {\bf 61}, 204 (2000). 

\bibitem{ricepile} K. Christensen, A. Corral, V. Frette, J. Feder, and 
  T. J{\o}ssang, Phys. Rev. Lett. {\bf 77}, 107 (1996).

\bibitem{notabtw}
        It is worth noticing that the original Bak, Tang and Wiesenfeld 
        sandpile model\cite{jenssen98} has deterministic 
        dynamics and does not belong to this universality class 
        (see Ref.\cite{bigfes}).

\bibitem{nota1dim}
        Preliminary results for the present universality class in 
        $d=1$ are reported in Ref.\cite{braz}. Noticeably, the CLG
        model in $d=1$ with sequential updating has been analytically
        solved by D. Dhar (private communication).

\end{thebibliography}
\end{document}